\documentclass[twoside,superscriptaddress,twocolumn,a4paper,floatfix,prl,aps,longbibliography]{revtex4-2}

\usepackage{xcolor}
\usepackage{graphicx}
\usepackage{bbold}
\usepackage[utf8x]{inputenc}
\usepackage[T1]{fontenc}
\usepackage{siunitx}
\usepackage{amstext}
\usepackage{hyperref}
\usepackage{amsfonts}
\usepackage{amsmath}
\usepackage{tabularx}
\usepackage{filecontents}
\usepackage{dsfont}
\usepackage{mathtools}
\usepackage{braket}
\usepackage{cases}

\newcommand{\outProd}[2]{|#1\rangle_{#2} \langle #1|}

\newcommand{\AB}[1]{\textcolor{black}{#1}}

\newcommand{\JK}[1]{\textcolor{black}{#1}}
\newcommand{\JKR}[1]{\textcolor{black}{#1}}

\begin{document}

\title{Resource-efficient entanglement detection in high-dimensional states via two-qubit witnesses}

\author{Josef Kadlec} 
\affiliation{Institute of Physics of the Academy of Sciences of the Czech Republic, Joint
Laboratory of Optics of Palacký University and Institute of Physics AS CR, 17.
listopadu 50a, 772 07 Olomouc, Czech Republic}
\affiliation{Palacký University in Olomouc, Faculty of Science, Joint Laboratory of Optics of Palacký University and Institute of Physics AS CR, 17. listopadu 12, 771 46 Olomouc, Czech Republic}

\author{Artur Barasi\'nski}
\affiliation{Institute of Theoretical Physics, University of Wroclaw, Plac Maxa Borna 9, 50-204 Wrocław, Poland}

\author{Karel Lemr} 
\affiliation{Palacký University in Olomouc, Faculty of Science, Joint Laboratory of Optics of Palacký University and Institute of Physics AS CR, 17. listopadu 12, 771 46 Olomouc, Czech Republic}

\begin{abstract}
This paper presents an efficient method for detecting entanglement in high-dimensional two-qudit states by mapping the Hilbert space onto the space of two qubits. This transformation enables the use of well-established two-qubit entanglement witnesses. The proposed approach is not restricted to any specific class of states, successfully identifies a vast majority of pure entangled states, and requires a number of measurements that does not increase with the dimensionality of the qudits. The method demonstrates solid sensitivity when applied to two notable classes of states—incomplete-permutation-symmetric states and random pure states mixed with white noise—and is shown to be feasible with current experimental techniques.
\end{abstract}

\date{\today}

\maketitle


\begin{figure}
\begin{center}
\includegraphics[scale=1]{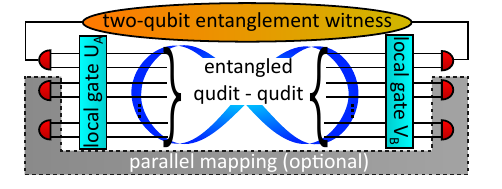}\\ 
\caption{Conceptual diagram of the proposed method. A two-qudit state is first subjected to a pair of local gates and then mapped onto the space of a two-qubit state. Standardized two-qubit entanglement witnesses can subsequently be used. To increase detection efficiency further, one can map the two-qudit space onto several two-qubit states detecting entanglement in them in parallel. \label{Fig_conceptual_diagram}}
\end{center}
\end{figure}

Natural quantum systems inherently store rich multidimensional quantum information, spanning across superpositions of more than two atomic \cite{Euler2023}, optical \cite{Ding2016} or mechanical modes \cite{Erhard2020}. 
The possibility of having entanglement between quantum systems with a large number of degrees of freedom opens interesting perspectives in quantum information science. 
Compared to qubit-based systems, entangled qudits can bolster Bell nonlocality \cite{Collins_PhysRevLett.88.040404} and mitigate detection loopholes \cite{Vertesi_PhysRevLett.104.060401}, facilitate high-capacity noise-resilient quantum cryptography \cite{Zahidy2024}, and a direct avenue for quantum simulations of complex molecular and physical systems \cite{Neeley_science2009, Blok_PhysRevX.11.021010, Choi2017}.
Crucially, universal quantum computation with qudits is viable in both circuit \cite{Luo2014, Chi2022} and measurement-based models \cite{Wei_PhysRevLett.106.070501}, requiring less resource overhead in quantum error correction and enhancing the execution of quantum algorithms \cite{Campbell_PhysRevLett.113.230501,Wang_PhysRevA.95.032312}. 
Recent experimental strides have demonstrated impressive progress in generating and manipulating high-dimensional entanglement \cite{Thew_PhysRevLett.93.010503,Brendel_PhysRevLett.82.2594,Jha_PhysRevLett.101.180405,MacLean_PhysRevLett.120.053601,Vaziri_PhysRevLett.89.240401,LeachScience329_2010,Krenn2017,Erhard2018,Howell_PhysRevLett.92.210403,SchaeffOptica2015,WangScience360_2018}.

This positions qudits as a compelling alternative to qubits in quantum technologies. However, entanglement detection in such systems remains a significant challenge. A common approach involves performing quantum state tomography followed by the application of an appropriate entanglement criterion to the reconstructed density matrix. From an experimental perspective, this method poses considerable technical difficulties, primarily due to the rapidly increasing resource demands. Specifically, for a bipartite system with local dimension $d$, full state tomography requires measurements in $d^4$ distinct projections \cite{Parisbook}.

As a more practical alternative, entanglement can be detected using witnesses, such as violations of Bell inequalities \cite{Barasiński_2024}, or by evaluating the fidelity of the unknown state with a maximally entangled state \cite{BourennanePRL92_2004}. These methods usually require significantly fewer measurements than full tomography. In particular, fidelity-based approaches can be implemented using structured measurement schemes, such as mutually unbiased bases (MUBs) \cite{Ivonovic_1981,Schwinger_1960,KrausPhysRevD35_1987} or equiangular measurements \cite{RenesJMP_2004,ZAUNER_2011}. For example, in the case of MUBs, the detection efficiency depends on the number and combinations of MUBs employed in the process \cite{MorelliPRL131_2023}. Unless all possible combinations are used, which again scales unfavorably as tomography, entanglement detection is not guaranteed even for pure states. 
Although various strategies exist for selecting a subset of $k$ MUBs from the complete set, constructing MUBs itself remains a nontrivial task. Moreover, the maximum number of mutually unbiased bases is still unknown for general Hilbert space dimensions.

In this paper, we propose a procedure that offers several key advantages over previously developed schemes. First, it is genuinely universal in the sense that it is applicable to arbitrary states without requiring prior knowledge. Second, the number of measurements remains constant and does not scale with the Hilbert space dimension, although the detection probability may decrease slightly with increasing dimension. Finally, our method reliably detects all pure entangled states and never misclassifies a separable state. The approach is based on a simple yet meaningful observation: any pure bipartite qudit state can be locally transformed into its Schmidt form, $|\psi\rangle = \sum_{j=0}^{r-1} \alpha_j |jj\rangle$, where $r$ stands for the Schmidt number. The state $|\psi\rangle$ is entangled if and only if $r\geq2$ \JKR{\cite{Terhal2000}}. This implies that any entangled state $|\psi\rangle$ can be mapped onto an entangled qubit-like state $|\tilde{\psi}\rangle_2 = N \sum_{j=0}^{1} \alpha_j |jj\rangle$, where $N$ denotes the standard normalization factor. Moreover, if $|\tilde{\psi}\rangle_2$ is entangled, it follows that $|\psi\rangle$ is also entangled. 
The central challenge lies in reliably identifying $|\tilde{\psi}\rangle_2$ as entangled among all possible two-qubit reductions. For an unknown state, the number of such two-qubit states scales as $d^4$. In this work, we show how this seemingly unfavorable scaling can be leveraged with high efficiency, enabling entanglement detection even for mixed bipartite qudit states.

\textit{Theoretical framework---}\AB{Let us consider a bipartite $d$-level (qudit) system described by the density matrix $\rho_d$ in $\mathbb{C}^d\otimes\mathbb{C}^d$. Our method begins with an optional local unitary transformation, $(U_A\otimes V_B)$, which does not affect the entanglement but can help improve detection in some cases (see Fig.~\ref{Fig_conceptual_diagram}).
Subsequently, the two-level subsystem is selected for each party using the operator $M = M_A P_{k} \otimes M_B P_{l}$. Here,  $M_A = \outProd{0}{A}+\outProd{1}{A}$ represents a superposition of projection operators acting on subsystem $A$ and $P_{k} = \sum_{i=0}^{d-1} |\pi^{(k)}_i\rangle\langle i|$ denotes a permutation operator with $\pi^{(k)}$ being the $k$-th permutation of $d$ levels. The same construction applies to subsystem $B$, with operators $M_B$ and $P_{l}$. 
The indices $k$ and $l$ are chosen randomly. In essence, the operator $M$ projects the bipartite qudit system into a two-qubit system, with the permutation operators introducing randomness in the selection of levels. The resulting two-qubit state can then be analyzed using standard tools for entanglement detection in two-qubit systems. In particular, the resulting two-qubit state given by 
\begin{equation}
\rho_2 = \frac{M \rho_d M^{\dagger}}{\text{Tr}\big[M \rho_d M^{\dagger}\big]},
\end{equation}
can be further written in a form
\begin{equation}
\rho_2 = \frac{1}{4}(\mathbb{1}\otimes \mathbb{1}
+\vec{a}\cdot\vec{\sigma} \otimes \mathbb{1}+
\mathbb{1}\otimes\vec{b}\cdot\vec{\sigma}
+ \sum_{m,n=1}^3 T_{mn}\sigma_m\otimes \sigma_n),
\end{equation}
where $\mathbb{1}$ stands for the identity operator, $\vec{\sigma}$ is the vector of standard Pauli matrices, and $\vec{a},\vec{b} \in \mathbb{R}^{3}$. The coefficients $T_{mn} = \text{Tr} \big[\rho_2 \sigma_m\otimes \sigma_n\big]$ form a real $3 \times 3$  matrix. To detect entanglement, we use the fully entangled fraction 
\begin{equation}\label{FEFw}
    \text{FEF\textsubscript{w}}(\rho_2) = 
    \frac{1}{2} \text{max}(0,\text{Tr} \sqrt{R}-1),
\end{equation}
where the real symmetric matrix $R = T^TT$. Entanglement is witnessed whenever $\text{FEF\textsubscript{w}}(\rho_2) >0$ \cite{Bartkiewicz2017}.
If entanglement is not detected in the first trial, the procedure can be repeated with different level combinations and/or local unitaries. We quantify the effectiveness of our method by the sensitivity, defined as the conditional probability of detecting entanglement given that the state is entangled \cite{Travnicek2024}.}


\AB{Note that FEF\textsubscript{w} was chosen as the entanglement witness due to its practical experimental accessibility. The correlation matrix $R$ can be obtained from the matrix $T$ using standard two-qubit quantum state tomography, which requires only 16 measurement settings per iteration \cite{Rehacek2004}. Alternatively, $R$ can be extracted from collective measurements \cite{Bovino2005, Bartkiewicz2013} on two copies of the state, reducing the requirement to just 10 projective measurements (see, e.g., \cite{Roik2022}). Importantly, this procedure does not scale with the system dimension $d\times d$, making it efficient even for high-dimensional systems. Nevertheless, FEF\textsubscript{w} can be substituted with any other two-qubit entanglement witness better suited to a specific experimental platform.}



\JK{To assess the experimental feasibility of our protocol across different physical implementations, we identify three key requirements that must be met: (i) the capability to generate genuine two-qudit entangled states, and (ii) the feasibility to filter arbitrary two-level subsystems within each qudit, which is essential for isolating the relevant qubit-like components, 
(iii) the capability to implement single-qudit local rotations.
In addition to these, there is an optional requirement in cases where collective measurements are desired, and (iv) the ability to perform Bell state projections on these subsystems .
Thanks to recent progress in quantum optics, all of these requirements—including the optional ones—are currently within reach. They are satisfied in systems based on multi-rail single-photon encoding \cite{Kok2007}, and even more convincingly in platforms using orbital angular momentum encoding. The latter has emerged as a particularly promising candidate for high-dimensional optical quantum communication, with strong experimental support from recent studies \cite{Krenn2014, Kysela2022, Leach2002, Mirhosseini2013, Yao2024}.}


\textit{Incomplete-Permutation-Symmetric states---}We demonstrate the application of our method on a special subgroup of the non-trivial family of mixed states, the incomplete-permutation-symmetric states (ICPS), which contains all bipartite states for finite dimension $d$, invariant under certain symmetry (see Ref. \cite{Barasinski2016, Barasinskipra95_2017} for details).
\AB{This particular choice of states is motivated by their relative complexity, which provides a meaningful challenge for entanglement detection, while still allowing us to effectively evaluate the protocol’s efficiency. Moreover, it enables us to demonstrate an experimentally practical strategy for mitigating the seemingly unfavorable scaling of detection efficiency with system dimension.}
The two-qudit ICPS states are defined as
\begin{equation}\label{ICPS}
\rho_{d}(\alpha,v) = v \ket{\psi_{\alpha}}\bra{\psi_{\alpha}} +
\frac{1-v}{d^2}\mathbb{1}_{d^2},
\end{equation}
where
\begin{equation}
\ket{\psi_{\alpha}} = \alpha \sum_{j = 0}^{r-2} \ket{jj} + \alpha_r \ket{(r-1)(r-1)},
\end{equation}
with $r$ denoting the Schmidt rank of $\ket{\psi_{\alpha}}$,  $0\leq \alpha\leq 1/\sqrt{r-1}$ and $\alpha_r = \sqrt{1-(r-1)\alpha^2}$. In other words, these states $\rho_d$ denote a mixture of pure states in their Schmidt form with white noise. Note that for $r=d$ and $\alpha=1/\sqrt{d}$ the state $\rho_d$ reduces to the isotropic states \cite{Horodeckipra59_1999}.

\AB{By analyzing all possible permutation operators $P_k$ and $P_l$ (see Supplemental Material), one finds that for the best combination of permutation operators without the use of local unitaries $\text{FEF\textsubscript{w}}(\rho_2) > 0$ when}
\begin{numcases}{v >}
   \dfrac{1}{1 + d^2 \alpha^2}  &$ \alpha \in (\frac{1}{\sqrt{r}},\frac{1}{\sqrt{r-1}})$
   \\
   \dfrac{1}{1 + d^2 \alpha \alpha_r} &$ \alpha \in (0,\frac{1}{\sqrt{r}})$
\end{numcases}
In particular, these conditions coincide with those that define the boundaries between state regimes characterized by various negative eigenvalues of partially transposed $\rho_d$ \cite{Barasinski2016}. Specifically, a single condition being satisfied corresponds to a single negative eigenvalue, whereas both conditions holding simultaneously indicate two negative eigenvalues. The above analytically proves that our method can detect all entangled ICPS states via the two-qubit FEF$_w$. The remaining question is its sensitivity. 

\begin{figure}
    \centering
   \includegraphics{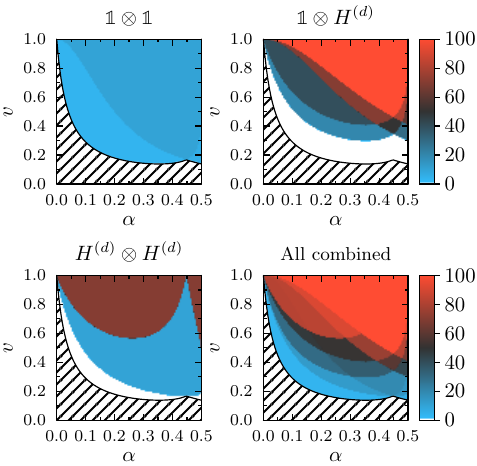}
    \caption{Colormap of the method sensitivity (\%) for various ICPS states parametrized by their visibility $v$ and balancing parameter $\alpha$ for $r = d = 5$. The operators on top of each subplot denote the operation performed on the respective qudits before selecting qubit subsystems. The final subplot (All combined) shows the success probability that \JKR{at least one} of the three preceding detections succeeds. Striped area represents separable states.
    \JKR{While this strategy requires at most 3 times the amout of measurement, it deliveres substantialy improved sensitivity. Note that even such increase in the number of measurements does not defeat the significant reduction in comparison to tomography.}}
    \label{ICCP_d5}
\end{figure}

\begin{table*}
\begin{center}
\begin{ruledtabular}
\begin{tabular}{lcccccccccccccc}
$d\times d$ &$3\times 3$& $3\times 3$  &$4\times 4$& $4\times 4$&$5\times 5$ &$5\times 5$& $6 \times 6 $&$6\times 6$& $7\times 7$  &$7\times 7$& $8\times 8$& $8\times 8$& $9\times 9$& $9\times 9$\\r &2& 3 &2&4&2 &5&2&6&2 &7&2&8&2&9\\ \hline 
single two-qubit choice:\\ \hline
(1)\quad$\mathbb{1} \otimes \mathbb{1}$& 10.8& 28.5& 2.8& 13.6& 1.0& 7.7& 0.4& 5.1& 0.2& 3.7& 0.1& 2.7&  0.1& 2.1 \\ 
(2)\quad$\mathbb{1} \otimes H^{(d)}$& 27.2& 68.2& 13.3& 52.8& 7.7& 50.1& 5.2& 40.4& 3.7& 42.5& 2.7& 37.6& 2.1& 36.7 \\ 
(3)\quad$H^{(d)} \otimes H^{(d)}$& 64.7& 55.2& 54.0& 41.3& 46.8& 32.6& 42.7& 27.3& 40.8& 24.4& 38.6& 21.8& 37.2& 19.6 \\
\JKR{(1) or (2) or (3)}& 73.4& 81.4& 59.4& 68.3& 50.3& 60.7& 45.2& 52.6& 42.6& 51.6& 40.1& 47.3& 38.4& 45.3 \\ 
 \hline
parallel approach:& & & & & & & & & & & & & &\\ \hline
(1)\quad $\mathbb{1} \otimes \mathbb{1}$& 10.8& 28.5& 5.6& 16.2& 2.0& 13.3& 1.3& 13.9& 0.7& 10.0& 0.5& 9.8& 0.3& 7.7 \\ 
(2)\quad $\mathbb{1} \otimes H^{(d)}$& 27.2& 68.2& 26.4& 66.0& 15.4& 62.6& 15.3& 63.7& 11.0& 60.9& 11.0& 63.1& 8.6& 59.8 \\
(3)\quad $H^{(d)} \otimes H^{(d)}$& 64.7& 55.2& 54.0& 41.3& 50.2& 41.0& 49.9& 43.9& 49.2& 41.6& 49.5& 44.1& 48.7& 42.1 \\
\JKR{(1) or (2) or (3)}& 73.4& 81.4& 64.4& 76.7& 56.6& 70.9& 55.7& 71.1& 53.5& 67.4& 53.8& 69.1& 52.1& 66.2 \\

\end{tabular}
\end{ruledtabular}
\end{center}\caption{Sensitivity (in percent) of our procedure on the ICPS class of states $\rho_d$ with various Schmidt ranks $r$ and dimension $d\times d$. The first column describes the operation on the two qudits before picking individual pairs of levels corresponding to the tested qubits. \label{TableAligned}}
\end{table*}

\AB{In the Supplemental Material, we derive an upper bound on the detection sensitivity, given by $\frac{2(r-1)r}{(d-1)^2d^2}$. Even in the most favorable case ($r=d$), this limits the detection probability to no more than $\dfrac{2}{(d-1)d}$, which decreases rapidly as $d$ increases. 
This behavior stems from the reduced degrees of freedom in the state $\ket{\psi_{\alpha}}$ imposed by the Schmidt decomposition, which confines the entanglement to pairs of identical levels across both subsystems.}

\AB{Our protocol provides a remedy to the aforementioned scaling issue. It significantly enhances the sensitivity by distributing the entanglement across a larger number of levels. This is accomplished by applying a local unitary transformation (LUT) to one or both subsystems prior to the action of the projective operators $P_k$ and $P_l$. }

\textit{Local Unitary Transformation---}As an example of LUT, let us consider the qudit Hadamard gate
\begin{equation}
H^{(d)} = \frac{1}{\sqrt{d}}\sum_{k=0}^{d-1}\sum_{l=0}^{d-1}\omega^{kl}\ket{k}\bra{l},
\end{equation}
where $\omega = \text{exp}{(\mathcal{I}2\pi/d)}$, with $\mathcal{I}$ denoting the imaginary unit, $k$ and $l$ representing basis elements. 
\AB{Exemplary results for $r=d=5$ are presented in Fig. \ref{ICCP_d5}. As shown, applying a single Hadamard gate ($\mathbb{1} \otimes H^{(d)}$) to one of the qudits significantly enhances the protocol’s sensitivity,  especially for highly entangled ICPS states, in some cases reaching perfect detection. However, certain weakly entangled ICPS states remain undetectable with this approach.
In contrast, applying Hadamard gates to both subsystems ($H^{(d)} \otimes H^{(d)}$) does not boost sensitivity as strongly as a single gate, but substantially increases the range of detectable entangled states compared to the single-gate approach}

 \begin{figure}
    \centering
    \includegraphics[width=1\hsize]{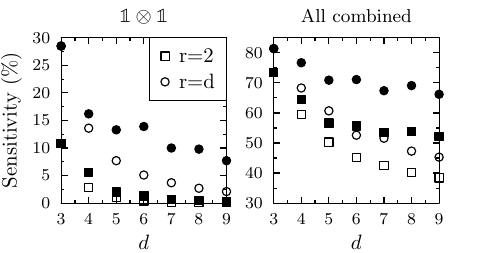}
    \caption{Graphical representation of sensitivity for the ICPS class of states $\rho_d$ with various dimension $d$ and $r$ either equal to 2 or the maximal value $d$. Empty symbols represent the single two-qubit choice, while the full symbols correspond to the parallel approach. The text on top denotes which unitary transformation was performed.}
    \label{Fig_Srank_2_max}
\end{figure}


To illustrate the impact of the Hadamard gate on sensitivity across a broader range of parameters, Table~\ref{TableAligned} presents typical results for the ICPS class, averaged over uniformly distributed $\alpha$ and $v$, for various values of  $d$ and $r$. The results demonstrate consistently high sensitivity, largely independent of the dimension of the system.
Notably, for $r=d$, the typical sensitivity appears to converge to around $36\,\%$ (at least for $d < 10$) when a single Hadamard gate is applied. It is significantly higher than in the case without the Hadamard gate ($\mathbb{1} \otimes \mathbb{1}$). 
The introduction of a second Hadamard gates to both subsystems nicely complements the single sided Hadamard gate for states of Schmidt rank r=2.
Overall, when all three detection scenarios are considered (All combined), the total sensitivity reaches approximately $45\,\%$ for $d=r=9$, representing a twenty-fold improvement over the case of no transformation.  It should be noted that, even in the most challenging case of $r=2$, the sensitivity remains relatively high, reaching about $38\,\%$ for $d=9$

\AB{The choice of the Hadamard gate is particularly appealing experimentally due to its ease of implementation. However, a natural question arises: can arbitrary LUTs enhance sensitivity in a similar way?
To investigate this, we analyzed the typical sensitivity of ICPS states under randomly sampled LUTs, averaging over $n=10^5$ uniformly distributed unitary matrices (see Fig.~2 in Supplemental Material). The results closely match those obtained with the Hadamard gate, indicating that the observed improvement in sensitivity is not unique to the Hadamard gate but reflects a more general effect.}


\begin{table*}
\begin{center}
\begin{ruledtabular} 
\begin{tabular}{cccccccc}
$d\times d$& $3\times 3$&$4\times 4$&$5\times 5$&$6\times 6 $ &$7\times 7 $ &$8\times 8 $ &$9\times 9 $  \\ \hline 
single two-qubit choice&&&&\\\hline
noise level  20\,\%& 96.0& 95.2& 95.0& 94.7& 94.6& 94.6& 94.6 \\ 
noise level  40\,\%& 75.9& 73.6& 72.5& 72.0& 71.7& 71.4& 71.3 \\ 
noise level 60\,\%& 28.6& 28.2& 28.1& 27.9& 27.8& 27.9& 27.9 \\  \hline
parallel approach &&&&\\\hline
noise level 20\,\%& 96.0& 99.8& 99.8& 100.0& 100.0& 100.0& 100.0 \\ 
noise level 40\,\%& 75.9& 94.0& 93.2& 98.2& 98.0& 99.4& 99.5 \\ 
noise level 60\,\%& 28.6& 50.8& 49.5& 64.4& 64.0& 74.5& 74.0 \\ 
\end{tabular}
\end{ruledtabular}
\end{center}\caption{ Sensitivity (in percent) of detecting  entanglement in a quasi-pure state for various dimensions and white noise levels \JKR{$1-v$}.\label{TableRnd}}
\end{table*}

\textit{Parallel detection---}
\AB{The probability of successfully detecting entanglement in two-qudit states can be further enhanced by simultaneously selecting multiple pairs of levels, as illustrated in Fig.~\ref{Fig_conceptual_diagram}. This approach maps a bipartite d-level system onto $\Big \lfloor \dfrac{d}{2} \Big \rfloor$ two-level subsystems, denoted $\rho_2^{(k)}$ for $k = 1,2 \dots, \Big \lfloor \dfrac{d}{2} \Big \rfloor$. Detecting entanglement in any one of these subsystems is sufficient to confirm entanglement in the original two-qudit state.
This parallel detection approach significantly enhances sensitivity, particularly for large $d$ (see Table~\ref{TableAligned}). Importantly, because all two-qubit measurements are performed in parallel, the overall number of required two-qudit state preparations does not increase. However, implementing parallel measurements across multiple subsystems poses additional experimental challenges.
When this method is combined with LUT applied to $\rho_d$, the typical sensitivity reaches approximately $66\%$ for $r=d=9$. Even in the most demanding scenario of minimal rank ($r=2$), the sensitivity remains high, reaching about $ 42\%$. Importantly, a further increase in $d$ does not appear to significantly reduce sensitivity, as suggested by the trend observed in Fig.~\ref{Fig_Srank_2_max}. This highlights the scalability and effectiveness of the combined strategy.
}

\textit{Random quasi-pure states---}In order to evaluate the performance of our procedure in realistic noisy conditions, we test it on randomly generated pure states affected by white noise
\begin{equation}
    \rho_{Q}(v) = v\ket{\psi_P}\bra{\psi_P}+\frac{1-v}{d^2} \mathbb{1}.
\end{equation}
\AB{The random pure states were generated by applying a random unitary matrix uniformly distributed according to Haar measure to a fixed pure state, as described in \cite{Singh2016}. For three representative noise levels (20, 40, and 60\,$\%$), we generated $10^5$ random states and subjected them to our entanglement detection procedure, simplified as no LUTs are needed, \JKR{since the set of random states is invariant under application of a fixed LUT.} 
The results in Table~\ref{TableRnd} indicate that for the most prominent high-purity states (noise below $ 20\,\%$) our procedure provides nearly perfect sensitivity, at least up to $d\leq 9$, in both single two-qubit and parallel detection approaches. While sensitivity decreases with increasing noise, it seems to be stable or even improves with dimension when using the parallel strategy. 
In the noisiest case (noise level of $60\,\%$), the parallel approach still yields a sensitivity of $74 \%$ for $d=9$. 
Remarkably, these results are achieved with no more than 16 measurement settings, independent of the system dimension $d\times d$, underscoring the protocol’s practical efficiency. They also confirm that ICPS states are among the most difficult to detect. In realistic settings, such as when the state is misaligned with the computational basis or subject to random local unitaries due to noisy quantum channels \JK{\cite{Knips2020, Jirakovaprapp16_2021},} our protocol remains effective without the need for local unitary adjustments, further simplifying its implementation.}



\textit{Conclusion---}The proposed protocol exhibits strong sensitivity in detecting entanglement in a wide variety of high-dimensional two-qudit states. \AB{Our protocol is based on two key requirements that can be met across different physical implementations. }
In contrast to previous work, this protocol is not limited to a particular class of states. Specifically, it successfully identifies \JKR{$45.3\,\%$ of ICPS states} with maximum Schmidt rank for qudit dimensions up to d= 9, and detects at least \JKR{$38.4\,\%$} ICPS states with Schmidt rank r=2\JKR{, see Table~\ref{TableAligned}}. The method is especially effective for quasi-pure random states, reliably identifying nearly all random entangled states with up to $20\,\%$ white noise. Even with white noise levels as high as $60\,\%$, the detection probability remains robust. Importantly, the number of measurement settings required does not increase with the dimension of the tested two-qudit state. Combined with its experimental feasibility, this protocol presents a promising approach for advancing quantum information processing using high-dimensional quantum systems.

\textit{Acknowledgments---}The authors thank CESNET for data management services and J. Cimrman for inspiration. 
The authors acknowledge support by the project of the Grant Agency of the Czech
Republic 25-17253S. All related programming is available via figshare \cite{kody}.

\bibliography{citace}

\section{Appendix}

\subsection{Efficient implementation via collective measurements}

\begin{figure}
    \centering
    \includegraphics{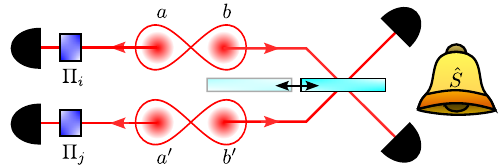}
    \caption{Scheme of the collective measurements on two identical copies of a two-qubit state. Parties $b$ and $b'$ are projected non-locally onto the Bell singlet state while parties $a$ and $a'$ are locally projected onto the minimal basis states \cite{Rehacek2004} defined by projectors $\Pi_i$ and $\Pi_j$. }
    \label{FIG_scheme}
\end{figure}

 \begin{figure}
    \centering
    \includegraphics{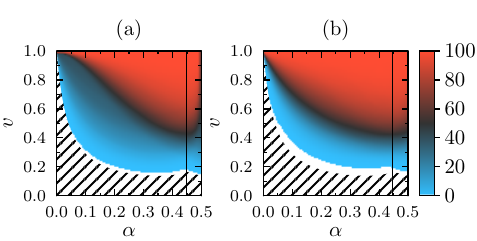}
    \caption{\JK{Colormap of the method sensitivity (\%) for various ICPS states under random unitary transformation for $r = d = 5$.  Plot (a) corresponds to local unitary performed on one subsystem, while in colormap (b) random local unitary was applied to both subsystems.  Striped area represents separable states.}}
    \label{ICCP_Unit_d5}
\end{figure}

Collective measurements, i.e. measurements on two or more identical copies of studied state, present a more efficient method for extracting properties of quantum states, compared to a standardly used measurements on only one copy. For example, given two copies of a single-qubit state, one can determine its purity using just one measurement setting \cite{Bartkiewicz2015}. To acquire the same information with only one copy of a given state, one would need to perform a full state tomography, consisting of at least 4 measurements in the so-called minimum basis set. Another example is the universal entanglement witness which requires only one measurement on four copies of a two-qubit state \cite{Bartkiewicz2017}, to determine whether the state is entangled or not, which is again a task requiring a full quantum state tomography consisting of at least $4\times4$ measurements considering only one copy.

Collective measurements on two copies of two-qubit states have recently earned an increased attention. 
One of the many possible scenarios has the same geometry as the entanglement swapping protocol, therefore can be readily applicable in near-future quantum communications. 
It is also experimentally more feasible than the universal entanglement witness. 
The second qubits from both copies are together subjected to a nonlocal projection  
\begin{equation}\label{Eq_S}
  \hat{S} = \mathbb{1} - 4\ket{\Psi^-}\bra{\Psi^-},
\end{equation}
with $\ket{\Psi^-} = (\ket{01}-\ket{10})/\sqrt{2}$  being the singlet Bell state,
while the first qubits are projected locally, see Fig.~(\ref{FIG_scheme}).

By having two identical copies of a two-qubit state $\hat\rho^{2\times 2}_{AB}$, where the subscript denotes the subsystem and superscript their dimension, the resulting state has a form of 
$\hat\rho^{2\times 2}_{ab}\otimes\hat\rho^{2\times 2}_{a'b'}$.  
One can draw the pairs of local projections imposed to $a$ and $a'$ from  many sets, but in practice one choose either eigenvectors of 
 the Pauli matrices $\sigma_i$ or the minimal basis set $\{\Pi_1,\Pi_2,\Pi_3,\Pi_4\}$ \cite{Rehacek2004} for both of $a$ and $a'$, while $b$ and $b'$ are projected nonlocaly using projector from Eq.~(\ref{Eq_S}).
 Due to the symmetry of the measurement scenario, the symmetric projections $L\otimes K$ and $K\otimes L$, where $L$ and $K$ are arbitrary projectors, give the same information, reducing significantly the number of measurements for the symmetric choice of bases.
 
There exist a quite straightforward method of extracting the individual matrix elements of $R$  from the obtained measurement results using the mean values of $\sigma$ 
\begin{equation}
R_{ij} = \text{Tr}\big[ 
\hat\rho^{(2\times 2)}_{ab}\otimes\hat\rho^{(2\times 2)}_{a'b'}
\hat{S}_{bb'}(\sigma_i\otimes\sigma_j)_{aa'}\big],
\end{equation}
and a simple transform between the results obtained using the minimum basis set $\Pi_i$ and the correlation matrix 
\begin{equation}
    R = (M^{-1})^T\pi M^{-1},
\end{equation}
where $M$ is the transformation matrix
\begin{equation}
    M = 
    \begin{pmatrix}
        1&s&s&s\\
        1&s&-s&-s\\
        1&-s&s&-s\\
        1&-s&-s&s
    \end{pmatrix}
\end{equation}
and the elements of $\pi$ correspond to the measured values of the projectors $\Pi_i\otimes \Pi_j$ 
\begin{equation}
\pi_{ij} = \text{Tr}\big[ 
\hat\rho^{(2\times 2)}_{ab}\otimes\hat\rho^{(2\times 2)}_{a'b'}
\hat{S}_{bb'}(\Pi_i\otimes\Pi_j)_{aa'}\big].
\end{equation}
Considering the aforementioned symmetry $\Pi_{ij} = \Pi_{ji}$, the total number of measurements amounts only to 10.
Several quantities can be calculated using the $R$ matrix, however, we will focus on the FEF based witness discussed in main text.

\subsection{Derivation of sensitivity for ICPS states}
\begin{table}
\begin{center}
\begin{ruledtabular} 
\begin{tabular}{cc}
 & Sensitivity\\ \hline
Eq.~(\ref{eq:FeF1a})& $\dfrac{2 (r-2) (r-1)}{(d-1)^2 d^2}$\\
Eq.~(\ref{eq:FeF1b})& $\dfrac{(r-1)}{(d-1)^2 d^2}$\\
Eqs.~(\ref{eq:FeF1a}) and (\ref{eq:FeF1b})& $\dfrac{2(r-1)r}{(d-1)^2 d^2}$
\end{tabular}
\end{ruledtabular}
\end{center}\caption{ Sensitivity for ICPS state of Schmidt rank $r$ and dimension $d\times d$ under the two possible outcomes determined by a specific permutation of levels [see Eqs.~(\ref{eq:FeF1a}) and (\ref{eq:FeF1b})]\label{TableProb}}
\end{table}

By analyzing all possible permutation operators $P_k$ and $P_l$, we find that the reduced state $\rho_2$ always takes one of four distinct forms, independent of the values of $d$ and $r$. These forms correspond to different scenarios based on the conditions:
\begin{eqnarray}
\langle 0|\pi^{(k)}_i\rangle &=& \langle 0|\pi^{(l)}_i\rangle=1\nonumber\\
\langle 1|\pi^{(k)}_j\rangle &=& \langle 1|\pi^{(l)}_j\rangle=1,
\label{condition}
\end{eqnarray}
where we assume, without loss of generality, that $i<j$. 
The scenarios are classified as follows: (i) The above conditions hold for $i,j=0,...,r-2$. (ii) The conditions hold for $i=0,...,r-2$ and $j=r-1$. (iii) Conditions \eqref{condition} is violated for $i,j=0,...,r-2$. (iv) Conditions \eqref{condition} is violated for $i=0,...,r-2$ and $j=r-1$.
This classification arises due to the inherent symmetry of ICPS states. Consequently, the function defining the FEF in Eq. \eqref{FEFw} also follows one of four possible expressions. However, only two of these can become positive and hence are relevant for entanglement detection, given by
\begin{numcases}{\text{FEF\textsubscript{w}}(\rho_2) =}
   \dfrac{3 d^2 v \alpha^2}{2+v (d^2 \alpha^2-2)} -1  & \label{eq:FeF1a}
   \\
   \dfrac{d^2 v \big[4 \alpha \alpha_r+1-(r-2) \alpha^2\big]}{4+v \big[(d^2-4)-(r-2) d^2 \alpha^2\big]}-1. & \label{eq:FeF1b}
\end{numcases}

To find the sensitivity of the protocol, one has to consider the number of permutation operators giving positive outcome of Eq.~(\ref{eq:FeF1a}) (the number of 2-combinations from the set of $r-1$ elements) and/or Eq.~(\ref{eq:FeF1b}) ($r-1$ possible choices) divided by the total number of operators $P_k$ and $P_l$. The results are presented in Table \ref{TableProb}.

\end{document}